# Observation of giant nonlinear valley Hall effect

Pan He[1,2†*], Min Zhang[1†], Jin Cao[3†], Jingru Li[1], Hao Liu[1], Jinfeng Zhai[1], Ruibo Wang[1], Cong Xiao[4*], Shengyuan A. Yang[3*] and Jian Shen[1,2,5,6,7,8*]

[1]*State Key Laboratory of Surface Physics and Institute for Nanoelectronic devices and Quantum computing, Fudan University, Shanghai 200433, China*

[2]*Hefei National Laboratory, Hefei 230088, China*

[3]*Institute of Applied Physics and Materials Engineering, Faculty of Science and Technology, University of Macau, Macau, China*

[4]*Interdisciplinary Center for Theoretical Physics and Information Sciences (ICTPIS), Fudan University, Shanghai 200433, China*

[5]*Department of Physics, Fudan University, Shanghai, China*

[6]*Shanghai Research Center for Quantum Sciences, Shanghai, China*

[7]*Zhangjiang Fudan International Innovation Center, Fudan University, Shanghai 201210, China*

[8]*Collaborative Innovation Center of Advanced Microstructures, Nanjing 210093, China*

[†]These authors contributed equally to this work.
[*]Correspondence to: hepan@fudan.edu.cn; congxiao@fudan.edu.cn; yangshengyuan@um.edu.mo; shenj5494@fudan.edu.cn

**Abstract**
**The valley Hall effect (VHE) holds great promise for valleytronic applications by leveraging the valley degree of freedom. To date, research on VHE has focused on its linear response to an applied current, leaving nonlinear valley responses undetected and nonlinear valleytronic devices undeveloped. Here, we report the experimental observation of a nonlinear VHE in a graphene-hBN moiré superlattice, evidenced by the generation of second-harmonic nonlocal voltages under AC currents. Remarkably, the nonlinear VHE has magnitude surpassing the linear VHE and is highly tunable via a gate voltage, which exhibits a pair of opposite peaks on the two sides of a Dirac gap. The nonlinear signal shows quadratic scaling with driving current and quartic scaling with local resistance, setting it apart from the linear counterpart. These experimental features are consistent with the theoretical picture of nonlocal transport mediated by nonlinear VHE and linear inverse VHE. We further reveal a nonlinear inverse VHE by observing the third- and fourth-harmonic nonlocal voltages. The nonlinear VHE provides a novel mechanism for valley manipulation and enables a novel valleytronic device—the valley rectifier—that converts AC charge current into DC valley current.**

# Main

The exploitation of charge and spin degrees of freedom of electrons has led to the booming fields of electronics and spintronics. In the past two decades, great efforts have been paid to utilizing the valley degree of freedom for information encoding and processing, giving birth to the field of valleytronics[1]. Valleys are local energy extrema of the electronic bands in momentum ($k$) space, which become well-defined degrees of freedom when they are well separated and intervalley scattering is weak. The generation of valley currents is at the heart of valleytronics, and the valley Hall effect (VHE) offers an efficient way to achieve this by purely electric means[2].

In VHE, electrons from opposite valleys are deflected to opposite directions, generating a valley current $j^v$ perpendicular to an applied electric field $E$ (Fig. 1a). The previously studied VHE is limited to the linear response regime, where $j_x^v = \sigma_H^v E_y$ with $\sigma_H^v$ the (linear) valley Hall conductivity. It arises from the valley-contrasting Berry curvature[2], which acts like a magnetic field in momentum space[3], and for nonmagnetic systems, its appearance requires the breaking of inversion symmetry. Experimentally, linear VHE has been detected in several 2D material systems, such as graphene and transition-metal dichalcogenides, via nonlocal measurement[4-8] or local measurement after optical pumping of valley carriers[9,10].

A frontier of current research is to discover new transport phenomena in the nonlinear regime. A range of nonlinear charge[11-18] and spin transport[19-23] effects have been studied. The whole picture would not be complete without a nonlinear VHE. Indeed, there have been theoretical proposals of nonlinear VHEs, with $j_x^v \propto E_y^2$ (Fig. 1b), e.g., from a nonlinear Drude mechanism in strained 2D transition-metal dichalcogenides[24], and more recently from a mechanism involving $E$ field correction to the Berry curvature[25]. However, the experimental demonstration of nonlinear VHE has not been achieved yet.

In this work, we report the observation of a giant nonlinear VHE in a graphene-hexagonal boron nitride (hBN) moiré superlattice. The nonlinear VHE is manifested by the second-harmonic generation under an AC driving current in the nonlocal measurement. We find this effect is highly gate tunable and can surpass the magnitude of its linear counterpart near Dirac points. The second harmonic nonlocal signal shows a quadratic scaling in driving current, a quartic scaling in local resistivity, and an opposite sign crossing the Dirac points, which are distinctive from linear signal but can be well explained as a consequence of nonlinear VHE followed by linear inverse VHE. Our analysis further reveals that the observed nonlinear VHE arises from an intriguing nonlinear zeroth-order extrinsic (ZOE) contribution. Furthermore, we also reveal a nonlinear inverse VHE, which combined with nonlinear VHE, produces the third- and fourth-harmonic nonlocal signals. Our work uncovers fundamental nonlinear valley transport effects in experiments and demonstrates their electric controllability, which lays a solid foundation for the emerging field of nonlinear valleytronics. These findings also directly lead to a novel device concept of valley rectifier, capable of generating DC valley currents from AC electric currents or electromagnetic radiation.

**Device characterizations and linear VHE**

We fabricated high-quality graphene-hBN moiré superlattices by stacking a monolayer graphene in between two hBN layers (Fig. 1c) via a dry transfer technique[26]. The crystalline axis of graphene was aligned with the top hBN and misaligned with the bottom hBN. The asymmetric stacking arrangement breaks the inversion symmetry of graphene[4,27], which opens a small gap at the Dirac points and induces significant Berry curvature as well as other band geometric quantities around the Dirac band edges[4]. Due to the small lattice mismatch between graphene and hBN, a long-period moiré superlattice is generated at the aligned interface[28-30]. It modifies the graphene band structure with the emergence of secondary Dirac points[29-31] below and above the two primary Dirac points. These features are manifested in the longitudinal resistance $R_{xx}$ and Hall resistance $R_{yx}$ versus the back gate voltage $V_g$ in Fig. 1d, obtained from a four-probe local measurement[4,27,29,30].

To detect VHE, we employed AC nonlocal measurements using the lock-in technique. This method enables the clear distinction between nonlinear and linear responses by measuring harmonic signals of different orders (Fig. 1c). The device and measurement geometry are shown in Fig. 1c. The graphene-hBN moiré superlattice is etched into a long strip geometry with width ~ 2 μm. Multiple pairs of contacts are made on the two sides of the strip to perform nonlocal measurements. For example, an AC current $I^\omega = I\sin\omega t$ is applied between contacts 1 and 2, then the VHE drives a valley current perpendicular to the applied current, which flows along the strip. The ultrahigh carrier mobility (μ ~ 400000 cm$^2$V$^{-1}$s$^{-1}$) enables long-range valley current transport over micrometer distances[4], so that the induced valley current can be converted into a nonlocal voltage $V_{NL}$ by the inverse VHE between some remote contacts 3 and 4. We repeated the measurement on 10 different devices, and the qualitative features reported below are common to all of them.

We first probe the linear VHE by measuring the first-harmonic nonlocal voltage $V_{NL}^{1\omega}$. The $V_{NL}^{1\omega}(V_g)$ curve exhibits sharp peaks in the narrow energy ranges at the primary Dirac points and the hole-side secondary Dirac points in Fig. 1e, where the existences of Berry curvature hot spot were found. The $V_{NL}^{1\omega}$ decays rapidly when the Fermi level moves away from these Dirac-point regions, consistent with previous report[4].

**Observation of nonlinear VHE**

To probe the nonlinear VHE, the second-harmonic nonlocal voltage $V_{NL}^{2\omega}$ was simultaneously measured. The physical picture is that the nonlinear VHE generates a valley current at frequency $2\omega$, which, by linear inverse VHE, produces the second-harmonic signal $V_{NL}^{2\omega}$ in the remote region (Fig. 1b). Remarkably, we observed a significant $V_{NL}^{2\omega}$ at the regions around the primary and hole-side secondary Dirac points (Fig. 1f). Importantly, different from $V_{NL}^{1\omega}$ which has a single peak and remains positive, $V_{NL}^{2\omega}$ exhibits a pair of opposite peaks and changes sign across each Dirac-point region (see Supplementary Fig. 1 for similar results obtained in other graphene-hBN moiré superlattice devices).

The $V_{NL}^{1\omega}$ signal scales linearly with the applied current amplitude $I$, as shown in Fig. 2a. We confirm that $V_{NL}^{2\omega}$ scales quadratically with $I$, as shown in Fig. 2b. Because of their distinct scaling relations, when simultaneously reversing the current and the voltage probe directions, the sign of $V_{NL}^{1\omega}$ remains unchanged, whereas the sign of $V_{NL}^{2\omega}$ should be inverted (Supplementary Fig. 2). This feature helps to eliminate background noise of $V_{NL}^{2\omega}$, which usually remains unchanged when the current and nonlocal probe directions are reversed. The heating or thermoelectric contribution to the signal can also be ruled out by swapping the driving and measuring contacts (Supplementary Fig. 3), since the thermoelectric contribution should flip sign under this swap whereas the VHE contribution should remain the same (Supplementary Note 1). After such careful analysis, we confirm that the $V_{NL}^{2\omega}$ can indeed be larger than $V_{NL}^{1\omega}$ at current amplitudes exceeding 2 μA for $V_g$ near the Dirac points (Fig. 2c, d) (see Supplementary Fig. 4 for similar results in other devices). This unexpected finding highlights a significant nonlinear effect contribution to nonlocal transport. Further confirmation of the second-order nonlinear nonlocal voltage was obtained through DC electric measurements (Supplementary Fig. 5), which revealed nonlinear *I-V* curves with the existence of large second-order terms.

**Scaling of second-order nonlocal resistance**

We next investigated the dependence of nonlocal transport on temperature ($T$). Figure 3a (3c) shows the gate voltage dependence of $V_{NL}^{1\omega}$ ($V_{NL}^{2\omega}$) at different $T$. Both $V_{NL}^{1\omega}$ and $V_{NL}^{2\omega}$ decease with rising $T$. In Fig. 3b (3d), the peak value of $V_{NL}^{1\omega}$ ($V_{NL}^{2\omega}$) around the secondary Dirac points is plotted as a function of $T$. One observes that both $V_{NL}^{1\omega}$ and $V_{NL}^{2\omega}$ exhibit an exponential decay in the range 60 K < $T$ < 180 K[4-7], as evidenced by the exponential fittings in Figs. 3b and 3d. As we will show below, the nonlocal resistance $R_{NL}^{1\omega}$ ($R_{NL}^{2\omega}$) scales with $(\rho_{xx})^n$ ($n$ is a positive integer), so the exponential decay of $V_{NL}^{1\omega}$ ($V_{NL}^{2\omega}$) may be attributed to the thermal activation behavior for $T > 60$ K, which leads to an exponential decrease of longitudinal resistivity $\rho_{xx}$ with rising $T$[27] (also see Supplementary Fig. 6). On the other hand, below 60 K, the data deviate from the exponential behavior. This can be explained by the fact that, at low temperatures, the system shifts to a regime dominated by hopping conduction between impurity states[4-7]. In addition, above 160 K, $V_{NL}^{1\omega}$ and $V_{NL}^{2\omega}$ become nearly undetectable, because of severe thermal smearing of Dirac band edges.

To explore the underlying physics of the observed nonlocal transport behavior, we perform a scaling analysis of the nonlocal resistance $R_{NL} = V_{NL}/I$. To do this, we fixed the gate voltage and analyzed how $R_{NL}$ scales with local resistivity $\rho_{xx}$ by varying the temperature in the thermal activation regime. It is known that the nonlocal transport from linear VHE combined with linear inverse VHE gives[32,33]:

$$R_{NL}^{1\omega} = \left(\frac{W}{2l_v}\right)(\sigma_H^v)^2 \rho_{xx}^3 \exp\left(-\frac{L}{l_v}\right), \qquad \text{Eq. (1)}$$

where $W$ is the width of the strip, $L$ is the length between the current-injection and voltage contacts, and $l_v$ is the valley diffusion length. When the intrinsic Berry-curvature-induced linear VHE is the dominant contribution, $\sigma_H^v \propto \rho_{xx}^0$[4]. Then, $R_{NL}^{1\omega}$ shows a $\rho_{xx}^3$ scaling, as observed in Fig. 4b, which is consistent with previous reports[4-7]. This cubic scaling is regarded as a hallmark of nonlocal transport mediated by linear direct and inverse VHEs, distinct from the classical nonlocal Ohmic (van der Pauw) resistance which scales as $\rho_{xx}^1$. It is also important to note that the result of Equation (1) is always positive, because $\sigma_H^v$ is squared in the expression. Here, one $\sigma_H^v$ is from the direct VHE process and the other is from the inverse process. This definite sign of the signal is consistent with the observation in Fig. 3a.

Remarkably, we find $R_{NL}^{2\omega}$ scales approximately as $\rho_{xx}^4$, as shown in Fig. 4a. This quartic scaling is also observed in other devices (Supplementary Fig. 7). It rules out possible contribution from nonlinear nonlocal Ohmic resistance, which should scale as $I\sigma_{nl}\rho_{xx}^3$, and thus at most as $\rho_{xx}^2$ ($\sigma_{nl}$ is the nonlinear charge Hall conductivity[12,15], whose highest scaling power is $\rho_{xx}^{-1}$). On the other hand, the quartic scaling is a distinctive characteristic of nonlocal transport mediated by nonlinear valley response. Consider the process of generation of valley current $j_x^v = \chi_{xyy}^v E_y^2$ by nonlinear VHE and its subsequent conversion to voltage signal by linear inverse VHE. Writing $j_x^v = (\chi_{xyy}^v E_y)E_y$, one may regard the valley current is from linear VHE with an effective conductivity $\chi_{xyy}^v E_y$ and hence may simply replace one $\sigma_H^v$ in Eq. (1) by $\chi_{xyy}^v E_y$. The resulting nonlocal resistance is given by

$$R_{NL}^{2\omega} \sim \left(\frac{I}{2l_v}\right) \chi_{xyy}^v \sigma_H^v \rho_{xx}^4 \exp\left(-\frac{L}{l_v}\right) \qquad \text{Eq. (2)}$$

Thus, one sees $R_{NL}^{2\omega}$ indeed shows a quartic scaling in $\rho_{xx}$, when $\chi^v \propto \rho_{xx}^0$ is contributed by the intrinsic[25] or ZOE nonlinear VHEs[34].

The simple argument leading to Eq. (2) reproduces the observed scaling. A more detailed derivation gives the following formula[35] (see Methods):

$$R_{NL}^{2\omega} = \left(\frac{I}{2\pi l_v}\right) \tilde{\chi}^v \sigma_H^v \rho_{xx}^4 \exp\left(-\frac{L}{l_v}\right) + \left(\frac{WI}{8l_v^2}\right)(\sigma_H^v)^2 \zeta^v \rho_{xx}^5 \exp\left(-\frac{2L}{l_v}\right), \qquad \text{Eq. (3)}$$

where $\tilde{\chi}^v = \chi_{xyy}^v + \chi_{xxx}^v$. Equation (3) has two terms. The first term is from nonlinear VHE combined linear inverse VHE. It is similar to Eq. (2) with a noticeable difference that not only the transverse ($\chi_{xyy}^v$) but also the longitudinal ($\chi_{xxx}^v$) nonlinear valley conductivities are involved. The second term, not present in Eq. (2), arises from linear VHE followed by nonlinear inverse VHE, with $\zeta^v$ being the nonlinear inverse valley Hall conductivity. One can show that the highest scaling of $\zeta^v$ is $\rho_{xx}^{-1}$, and therefore this term also scales as $\rho_{xx}^4$.

**Theoretical understandings**

To determine which one of the two terms in Eq. (3) accounts for the observation, we made use of their distinction in the exponential decay factor. Specifically, we measured the nonlocal voltages at different probing distance ($L$) from the current channel. From the result of $R_{NL}^{1\omega}$ (Fig. 4d), we deduce that the valley diffusion length $l_v$ is around 1 micron using Eq. (1). Next, focusing on $R_{NL}^{2\omega}$, we observe that it also decays exponentially with $L$ (Fig. 4c). Fitting the data with the factor $\exp\left(-\alpha \frac{L}{l_v}\right)$, we find that $\alpha \simeq 1$. Therefore, the first term in Eq. (3), which manifests the nonlinear VHE followed by linear inverse VHE, dominates over the second term (which is more rapidly damped with $L$) in the observed signal.

The nonlinear VHE physics also explains the sign change of $V_{NL}^{2\omega}$ as the gate voltage passes Dirac points, which is especially salient around the primary Dirac points (Fig. 3c). Different from the linear signal, according to Eq. (2) or Eq. (3), the sign of $V_{NL}^{2\omega}$ is not definite. Instead, it is determined by that of $\tilde{\chi}^v \sigma_H^v$ (here we focus on the first term of Eq. (3) which dominates). Around a primary Dirac point, there are two Dirac bands with approximate particle-hole symmetry. We find that for such a case, regardless of details of scatterers, both the intrinsic and the ZOE contributions to $\tilde{\chi}^v$ ($\sigma_H^v$) are antisymmetric (symmetric) with respect to carrier density or chemical potential (Supplementary Note 2). As such, $\tilde{\chi}^v \sigma_H^v$ must be an odd function across the Dirac point. This explains the observed antisymmetric pattern in $V_{NL}^{2\omega}$.

It is worth noting that the nonlinear VHE itself does not require broken inversion symmetry, but the above analysis show that its nonlocal measurement involves the linear inverse VHE, which does require broken inversion symmetry. Consequently, $V_{NL}^{2\omega}$ is expected to vanish in graphene-hBN heterostructures without interface alignment, where inversion symmetry is preserved. Our experiments on such control samples indeed confirmed this by showing much smaller magnitude (or absence) of $R_{NL}^{1\omega}$ and $R_{NL}^{2\omega}$ (Supplementary Fig. 8). For a new application, the nonlinear VHE enables a valley rectifier that converts AC electric currents into DC valley currents. The DC valley current can be detected as a DC nonlocal voltage (Supplementary Fig. 9).

Having shown that nonlinear VHE offers a consistent explanation of all the key features observed for the second harmonic signal, we further investigate the possible physical mechanisms for this nonlinear VHE in our system. First, we note that the nonlinear Drude mechanism from the Fermi pocket anisotropy gives a scaling $\tilde{\chi}^v \propto \rho_{xx}^{-2}$[24], leading to $R_{NL}^{2\omega} \propto \rho_{xx}^2$, which is contradictory to the observation. Hence, it can be excluded. Second, as mentioned, nonlinear valley conductivity with $\tilde{\chi}^v \propto \rho_{xx}^0$ may arise from intrinsic[25] or ZOE[34] mechanisms. The intrinsic nonlinear VHE originates from the electric field correction to Berry curvature[11], quantified by Berry connection polarizability[36]. Notably, it requires the breaking of $C_3$ symmetry, which is not fulfilled in our system. Hence, the intrinsic mechanism is also excluded. The only candidate left is thus the ZOE mechanism. This type of nonlinear VHE can be connected to the ZOE nonlinear charge transport for each valley, which involves, e.g., composition of side jump and skew scattering processes, their field corrections, and their interplay with Berry-curvature anomalous velocity[34]. This is the first time that nonlinear ZOE transport is detected in nonmagnetic systems and in valley physics.

**Evidence of nonlinear inverse VHE**

In Eq. (3), the second term involves a nonlinear inverse VHE, but as we have seen, this term is overwhelmed by the first term in $V_{NL}^{2\omega}$. Then, how can we probe this nonlinear inverse VHE? Fortunately, it enters the third- and fourth-harmonic signals, when combined with nonlinear direct VHE (Fig. 5a). Indeed, we have observed sizable nonlocal voltages $V_{NL}^{3\omega}$ and $V_{NL}^{4\omega}$, which exhibit cubic and quartic scaling of driving current, respectively (Fig. 5b and 5c). The theoretical analysis based the picture of nonlinear direct and inverse VHEs predicted that $R_{NL}^{3\omega} = V_{NL}^{3\omega}/I$ should $\propto \rho_{xx}^5$ [35], which also agrees with our observation in Fig. 5d.

**Conclusion and Outlook**

In summary, we have presented the experimental detection of nonlinear VHE in monolayer graphene-hBN moiré superlattices. This effect manifests in the second-harmonic generation of a nonlocal voltage under an AC driving current. We observe a large nonlinear signal at both the primary and secondary Dirac points, surpassing the linear counterpart in magnitude. Furthermore, we reported evidence of nonlinear inverse VHE through the observation of third- and fourth-harmonic nonlocal signals. Our work establishes a novel fundamental effect of nonlinear transport and offers a novel approach to generate valley current. The findings here directly enable a valley rectifier device, which can convert AC electric signals or electromagnetic waves into DC valley current. This holds great potential for valleytronic applications.

As an outlook, the nonlinear VHE here can be further manipulated by tailoring the strength of inversion symmetry breaking in monolayer graphene through substrate engineerings[37,38] or adjusting interlayer twist angles[39]. Additionally, this effect is anticipated in bilayer and few-layer graphene as well as transition metal dichalcogenides[40]. Electrical control of nonlinear VHE, achievable by applying an interlayer gate bias to tune the bandgap and Berry curvature, is a particularly intriguing avenue in these materials. Furthermore, temperature gradients can induce nonlinear effects analogous to electric fields, such as the nonlinear valley Nernst effect[41] and nonlinear valley thermal Hall effect, which are waiting for future explorations.

**Methods**

**Device fabrication**

The hBN/graphene/hBN heterostructures were fabricated using a dry-transfer technique. First, hBN (HQ Graphene) and graphene (HOPG, HQ Graphene) were mechanically exfoliated onto separate $Si/SiO_2$ (285 nm) substrates. Suitable hBN thin films (20-40 nm thick) and monolayer graphene flakes were identified using optical microscopy. Monolayer graphene was further confirmed by Raman spectroscopy. Next, the dry transfer process with polycarbonate (PC) film mounted on a thick polydimethylsiloxane (PDMS) stamp was employed to pick up the top hBN layer. The graphene layer was then transferred onto the top hBN using the same method, with careful alignment of their crystal axes (typically by the straight edges) to form the moiré superlattice. Then, the bottom hBN layer was pick up to complete the heterostructure. Finally, the whole stack was positioned onto a highly-doped silicon wafer at the PC melting point of 180 °C. The full width at half maximum (FWHM) of the Raman 2D-band was used to estimate the twist angle between hBN and graphene. For the Hall bar devices used in nonlocal measurements, a two-step process involving electron beam lithography and reactive ion etching was employed. Electron beam resists MMA EL6 and PMMA A4 were spin-coated onto the sample for the etch mask writing. After etching by $CHF_3/O_2$ plasma and cleaning the resist, electron beam lithography was used to define the pattern of electrodes on a freshly spin-coated resist layer. The electrodes of Cr/Au (3 nm/90 nm) metal layers were deposited onto the defined areas using thermal evaporation to form the final device.

**The nonlocal electrical measurements**

The nonlocal electric measurements were performed in a Physical Properties Measurement System (PPMS, Quantum Design). All nonlocal measurements were carried out at zero magnetic field in this work. For AC nonlocal measurements, a sinusoidal current $I^\omega = I\sin\omega t$ with a frequency $\omega/2\pi$=17.777 Hz was applied to one pair of Hall arms (perpendicular to the longitudinal stripe of the device) using a source meter (Keithley 6221). The different harmonic nonlocal voltages were simultaneously measured by multiple lock-in amplifiers (Stanford Research SR830) from another pair of Hall arms in the device. The phases of lock-in amplifiers were set to 0° and 90° for the odd and even harmonic voltage measurements, respectively. For DC nonlocal measurements, the Keithley 2400 source meter and Keithley 2182A voltmeter were used. For rectification measurements, the Keithley 6221 source meter and Keithley 2182A voltmeter were employed. A sub-femtoamp source meter (Keithley 6430) was used for back-gate voltage application and leakage current measurements.

**Derivation of the valley mediated nonlinear nonlocal resistance**

This section details the frequency-dependent nonlocal resistance at the double-, third-, and fourth-harmonic channels[35]. In establishing a nonlocal signal by charge-valley interconversion, what lies between the VHE happening in the input-current region and the inverse VHE in the voltage probe region is the diffusion of valley density imbalance, which can be described by the one-dimensional diffusion equation:

$$D\partial_x^2 n^v(x) - \frac{n^v(x)}{\tau_v} = \Xi(x). \tag{1}$$

Here $n^v(x)$ is the valley imbalance averaged over $y$ direction, $D$ is the diffusion constant, $\tau_v$ is the valley relaxation time, and

$$\Xi(x) = \frac{1}{W}\int_{-W/2}^{W/2} dy \partial_a(\chi_{abc}^v E_b E_c) \tag{2}$$

is the generation rate of the valley imbalance from nonlinear valley current. In steady state, the distribution of the electric field obeys the 2D Laplace equation $\nabla^2 \phi = 0$ subjected to a boundary condition $j_y(x, y = \pm W/2) = I\delta(x)$, which reflects the applied driving current at $x = 0$ (see Fig. 1 in the main text). The solution is given by:

$$E_y = -\frac{2\rho I}{W} \frac{\sinh(\pi x/W) \sin(\pi y/W)}{\cosh(2\pi x/W) + \cos(2\pi y/W)} \tag{3}$$

and

$$E_x = \frac{2\rho I}{W} \frac{\cosh(\pi x/W) \cos(\pi y/W)}{\cosh(2\pi x/W) + \cos(2\pi y/W)}. \tag{4}$$

By noting that $E_x(x, y) = -E_x(x, -y)$ and $E_y(x, y) = E_y(x, -y)$, one can verify that only $\chi^v_{xyy}$ and $\chi^v_{xxx}$ are relevant to nonlocal transport. Their contribution to the valley generation rate is

$$\Xi(x) = -\frac{2\pi}{W} (\chi^v_{xyy} + \chi^v_{xxx}) \left(\frac{\rho I}{W}\right)^2 \frac{\cosh(2\pi x/W)}{\sinh^2(2\pi x/W)}. \tag{5}$$

Solving the diffusion equation yields the nonlinear valley imbalance

$$n^{v;NL}(x) = \frac{\tau_v \rho^2 I^2}{\pi W l_v^2} (\chi^v_{xyy} + \chi^v_{xxx}) \exp\left(-\frac{x}{l_v}\right). \tag{6}$$

It is interesting to note that in addition to the transverse response $\chi^v_{xyy}$, the longitudinal response $\chi^v_{xxx}$ also contributes. This differs from the linear case, in which the same kind of derivation shows only the Hall response is relevant to the linear valley imbalance:

$$n^{v;L}(x) = \frac{\tau_v \rho I}{2 l_v^2} \sigma^v_H \exp\left(-\frac{x}{l_v}\right). \tag{7}$$

Equation (6) yields a driving force $F^{NL}_x = -\partial_x n^{v;NL}/2\nu$ for inverse VHE ($\nu$ is the density of states of one valley), which generates a charge current that is quadratic in $I$:

$$j_y^{(2,1)} = \frac{1}{2} \sigma^v_H F^{NL}_x. \tag{8}$$

Using the relation $D = 1/(2\nu\rho)$, a second-harmonic nonlocal resistance can be obtained:

$$R^{(2;1)}_{NL}(x) = \frac{\rho W j_y^{(2,1)}}{I} = \left(\frac{I}{2\pi l_v}\right) (\chi^v_{xyy} + \chi^v_{xxx}) \sigma^v_H \rho^4 \exp\left(-\frac{x}{l_v}\right). \tag{9}$$

On the other hand, the valley-to-charge conversion can also be realized by a nonlinear inverse VHE

$$j_y^{(2,2)} = \frac{1}{2} \zeta^v (F^L_x)^2, \tag{10}$$

where $F^L_x = -\partial_x n^{v;L}/2\nu$ is the valley imbalance from linear direct VHE, and the nonlinear inverse valley Hall conductivity $\zeta^v$ is equal to the nonlinear charge Hall conductivity of the valley system. $j_y^{(2,2)}$ gives another contribution to the second-harmonic nonlocal resistance:

$$R^{(2;2)}_{NL}(x) = \frac{\rho W j_y^{(2,2)}}{I} = \left(\frac{WI}{8 l_v^2}\right) (\sigma^v_H)^2 \zeta^v \rho^5 \exp\left(-\frac{2x}{l_v}\right). \tag{11}$$

The sum of Eqs. (9) and (11) is Eq. (3) in the main text.

The third-harmonic charge current at position $x$ arises via nonlinear inverse VHE, which is given by $j_y^{(3)} = \frac{1}{2} \zeta^v F^L_x F^{NL}_x$, with the composition of $F^L_x$ and $F^{NL}_x$ as driving force. Additionally, through the interplay between the nonlinear VHE and the nonlinear inverse VHE, a fourth-harmonic charge current is generated: $j_y^{(4)} = \frac{1}{8}\zeta^v(F_x^{NL})^2$. The two charge currents lead to nonlocal resistance at $3\omega$ and $4\omega$ channels, given respectively by

$$R_{NL}^{(3)}(x) = \left(\frac{I^2}{4\pi l_v^2}\right)\zeta^v \sigma_H^v (\chi_{xyy}^v + \chi_{xxx}^v)\rho^6 \exp\left(-\frac{2x}{l_v}\right) \quad (12)$$

and

$$R_{NL}^{(4)}(x) = \left(\frac{I^3}{8\pi^2 W l_v^2}\right)\zeta^v (\chi_{xyy}^v + \chi_{xxx}^v)^2 \rho^7 \exp\left(-\frac{2x}{l_v}\right). \quad (13)$$

Suppose $\sigma_H^v, \chi^v \propto \rho^0$ are dominated by intrinsic[2,11,25] or zeroth-order extrinsic mechanisms[34], and $\zeta^v$ scales as $\rho^{-1}$, they scale as $R_{NL}^{(3)} \propto \rho^5$ and $R_{NL}^{(4)} \propto \rho^6$, respectively.

**Acknowledgments**

P.H. was sponsored by National Key Research and Development Program of China (grant no. 2022YFA1403300 and 2020YFA0308800), National Natural Science Foundation of China (grant no.12174063 and U23A2071), Natural Science Foundation of Shanghai (23ZR1403600) and the start-up funding from Fudan University. This work was supported by Innovation Program for Quantum Science and Technology grant 2024ZD0300103. C.X. was sponsored by the start-up funding from Fudan University. S.A.Y. was supported by The Science and Technology Development Fund of Macau SAR (FDCT 0066/2024/RIA).

**Author contributions**

P.H. and J.S. planed the study. P.H. and M.Z. performed the transport measurements and the data analysis. M.Z., J.L., H.L., J.Z. and R.W. fabricated devices. J.C., C.X. and S.A.Y. performed theoretical studies. P.H., M.Z., J.C., C.X., S.A.Y. and J.S. wrote the manuscript. All authors commented the manuscript.

**Competing interests** The authors declare no competing interests.

**Data availability**

The data that support the findings of this study are available within the paper and the Supplementary Information. Other relevant data are available from the corresponding authors upon reasonable request. Source data are provided with this paper.

**Code availability**

The codes that support this study are available from the corresponding author upon reasonable request.

# Figures

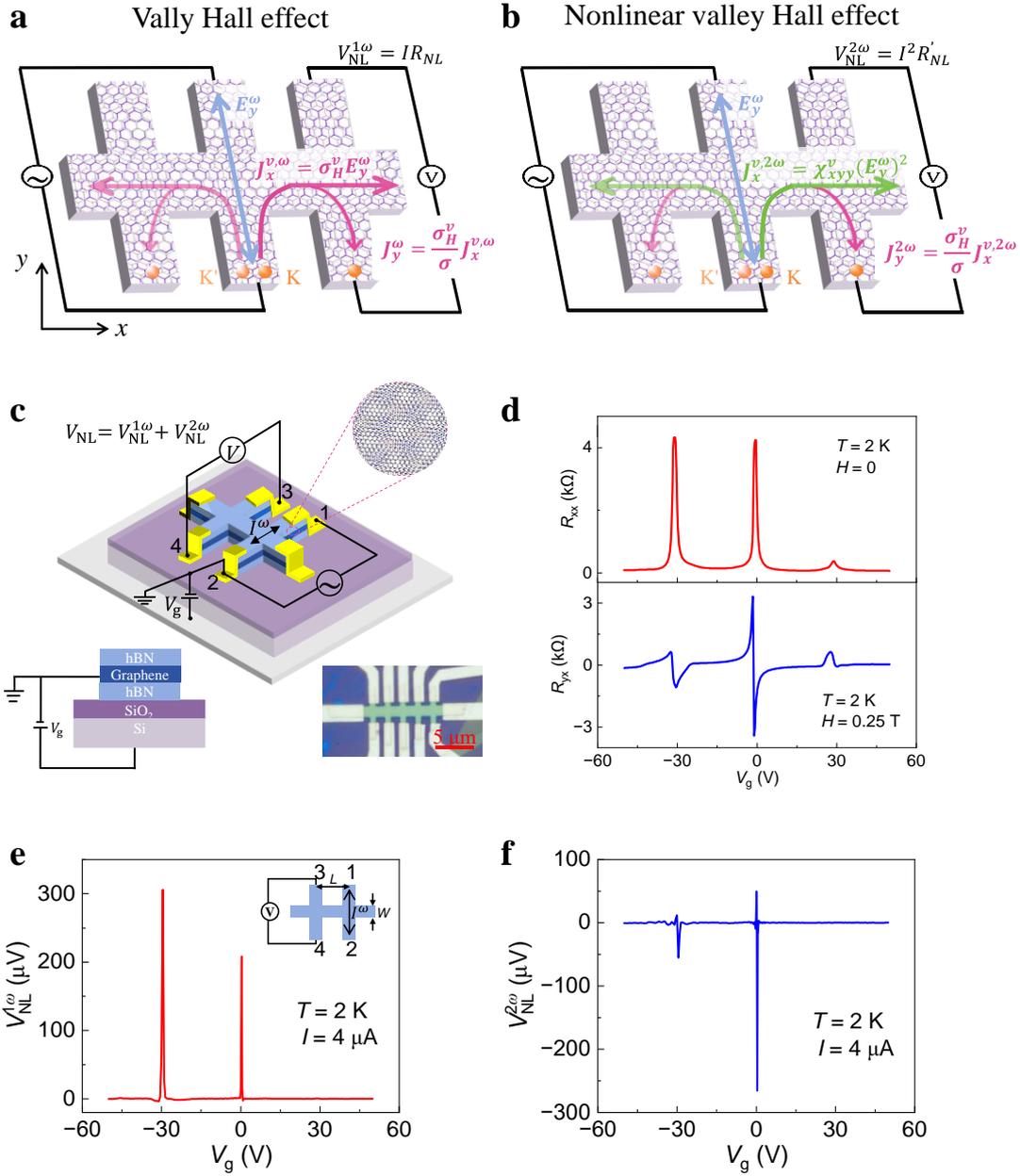

**Fig. 1| Schematic of the valley Hall effect and nonlinear valley Hall effect as well as their nonlocal measurements in a graphene-hBN moiré superlattice. a**, Schematic of the valley Hall effect (VHE) and the nonlocal measurement via the inverse VHE. An AC electric field $E_y^\omega$ with frequency $\omega$ is applied in the middle-pair of Hall arms, which generates a pure valley current $J_x^{v,\omega}$ proportional to $E_y^\omega$ in the transverse direction by the VHE with a conductivity $\sigma_H^v$. The valley current is further converted into a charge current by the inverse VHE at the left- and right-pair of Hall arms, generating detectable nonlocal voltages. **b**, Schematic of the nonlinear VHE and the nonlocal measurement via the inverse VHE. A nonlinear valley current $J_x^{v,2\omega}$ proportional to the $(E_y^\omega)^2$ is generated through a nonlinear valley Hall conductivity $\chi_{xyy}^v$, which is subsequently converted into a measurable second-harmonic nonlocal voltage in the remote Hall arms. The linear (**a**) and nonlinear (**b**) responses are represented by the pink and green arrows, respectively. **c**, Schematic of the device structure of hBN-encapsulated monolayer graphene and the electric

harmonic measurements, where the first- and second-harmonic nonlocal voltages were measured simultaneously upon the application of an AC current. Inset shows the side view of sample structure and an optical image of a Hall bar device with a channel width $W = 2$ μm and a length $L = 2$ μm (of the closest arms). **d,** The longitudinal resistance $R_{xx}$ (top panel) and Hall resistance $R_{yx}$ (bottom panel) as functions of back gate voltage $V_g$. The secondary Dirac points emerge at $V_g$ = -31 V and 29 V (Device 1). **e, f,** The first-harmonic nonlocal voltage $V_{NL}^{1\omega}$ (**e**) and the second-harmonic nonlocal voltage $V_{NL}^{2\omega}$ (**f**) as functions of $V_g$ at $I = 4$ μA. All the data were measured under zero magnetic field except the $R_{yx}$ in (**d**).

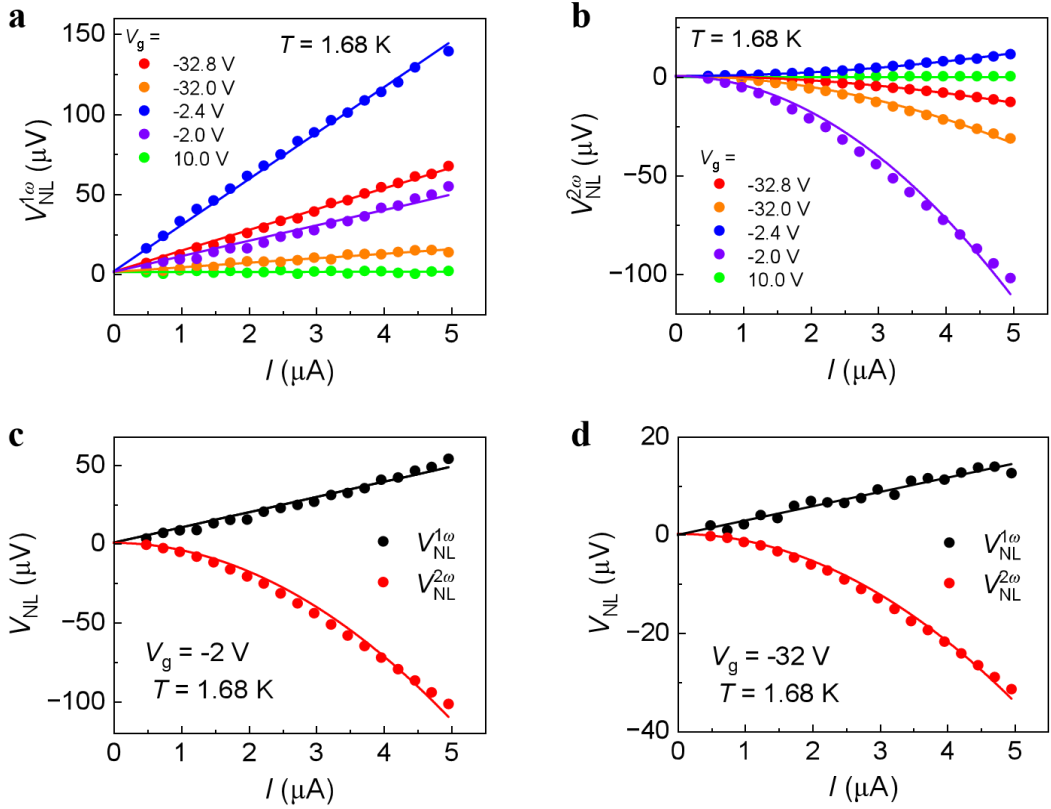

**Fig. 2| The first- and second-harmonic voltages in nonlocal measurements (Device 1). a,** The first-harmonic nonlocal voltage $V_{NL}^{1\omega}$ as a function of current amplitude $I$ for several $V_g$. The solid lines are linear fittings to the data. **b,** The second-harmonic nonlocal voltage $V_{NL}^{2\omega}$ versus $I$ for several $V_g$. The solid lines are quadratic fittings to the data. **c, d,** Two sets of data from (**a, b**) are replotted here to show that the $V_{NL}^{2\omega}$ can be larger than the $V_{NL}^{1\omega}$ at a $V_g$ around the primary (**c**) and the secondary Dirac points (**d**), respectively. The measurements were performed at zero magnetic field.

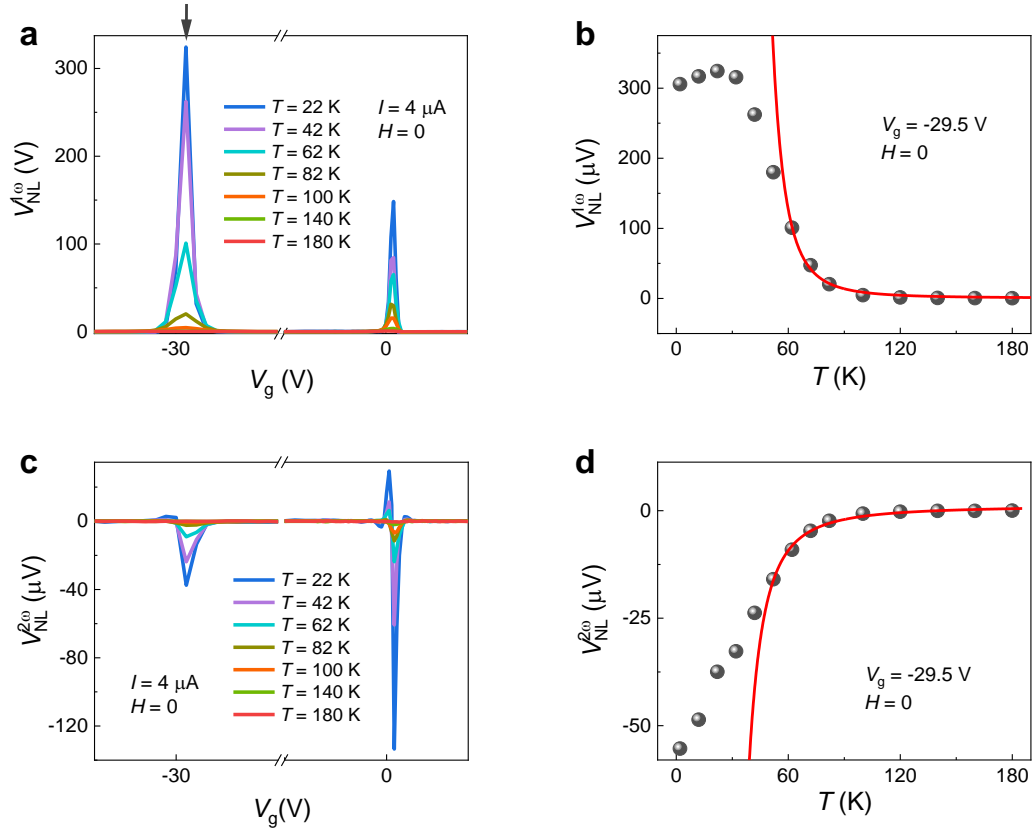

**Fig. 3| Temperature dependence of the first- and second-harmonic nonlocal voltages (Device 1). a**, The $V_{NL}^{1\omega}$ as a function of $V_g$ measured at several temperatures. **b**, The $V_{NL}^{1\omega}$ peak, as marked by the arrow in (**a**), as a function of temperature. **c**, The $V_{NL}^{2\omega}$ as a function of $V_g$ measured at several temperatures. **d**, The $V_{NL}^{2\omega}$ peak as a function of temperature. The red lines in (**b**, **d**) show exponential fittings in the thermal activation regime of 60 K < $T$ < 180 K.

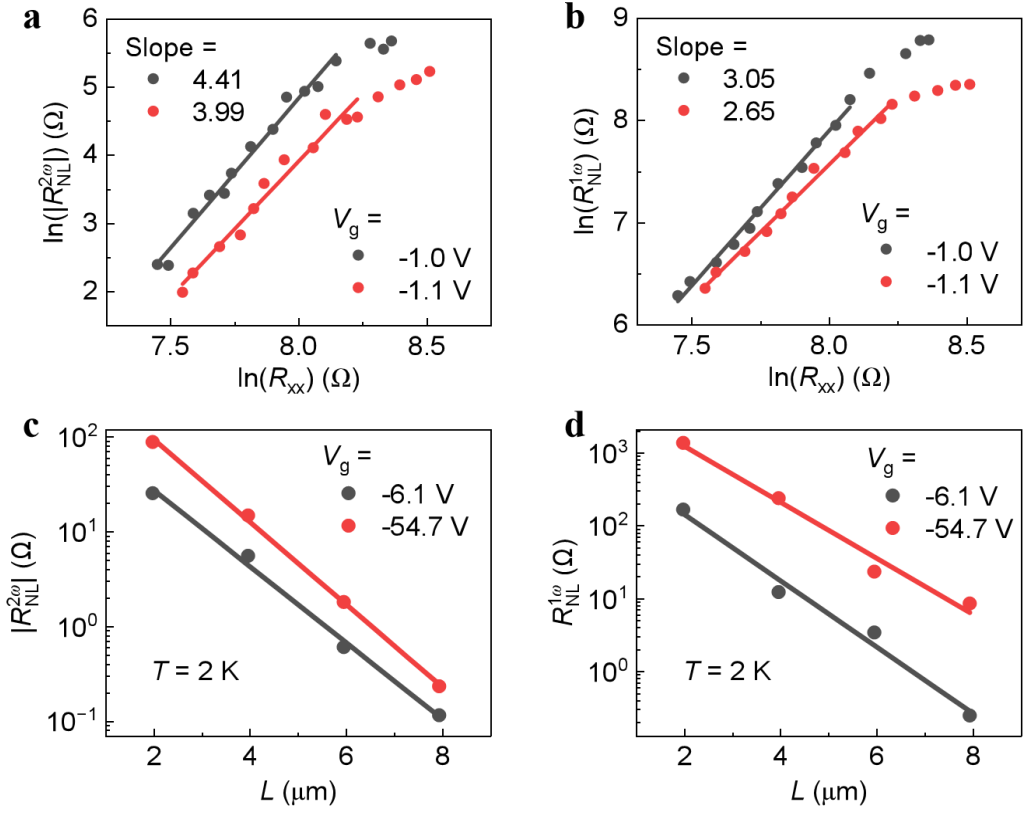

**Fig. 4| The scaling relation between the nonlocal resistance $R_{NL}$ and the local one $R_{xx}$. a,b**, The $\ln R_{NL}^{2\omega}$ (**a**) and $\ln R_{NL}^{1\omega}$ (**b**) are plotted against $\ln R_{xx}$ at two different $V_g$ in Device 2. The linear fittings to the data in the thermal activation regime are shown as the solid lines. The fitting slops are indicated. **c, d**, The $R_{NL}^{2\omega}$ (**c**) and $R_{NL}^{1\omega}$ (**d**) as a function of probing distance $L$ for a $V_g$ near the primary Dirac points and secondary Dirac points, respectively, in Device 1.

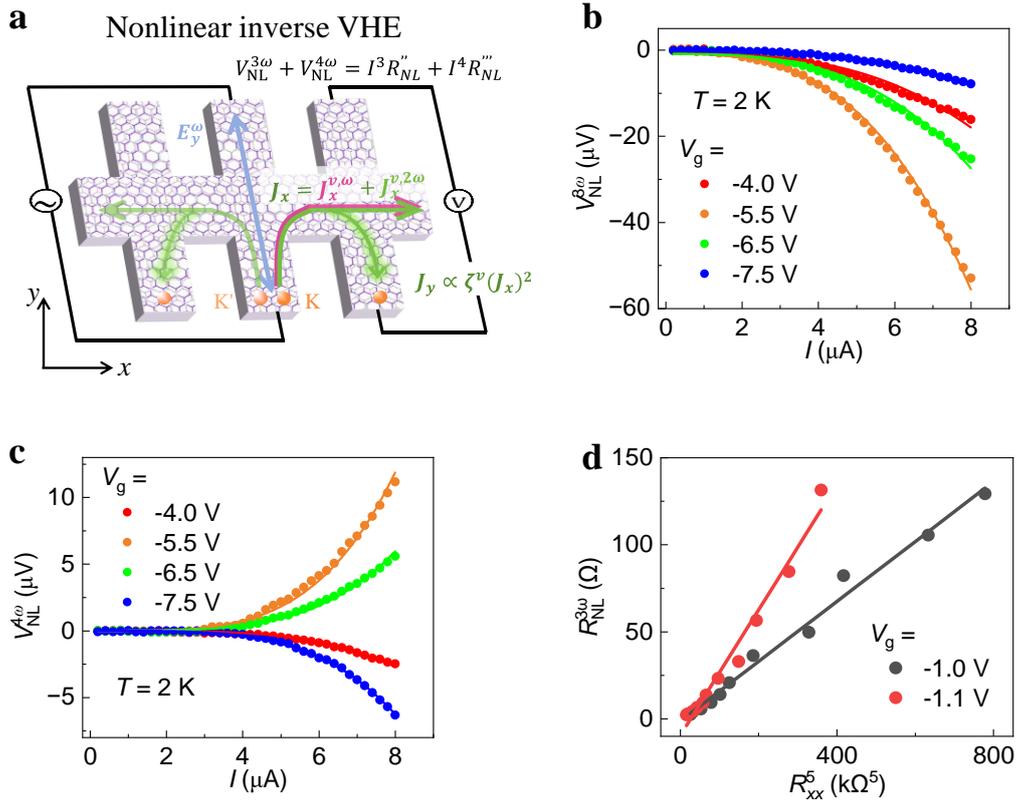

**Fig. 5| The nonlinear inverse VHE and the third-harmonic and fourth-harmonic nonlocal voltages. a**, Schematic of the nonlinear VHE and the nonlinear inverse VHE (highlighted with edge glow) in the nonlocal measurement setup. **b**, **c**, The third-harmonic nonlocal voltage $V_{NL}^{3\omega}$ (**b**) and fourth-harmonic nonlocal voltage $V_{NL}^{4\omega}$ (**c**) as a function of $I$ for several representative $V_g$ in Device 3. The solid lines are cubic fittings (**b**) and quartic fittings (**c**) to the data. **d**, The $R_{NL}^{3\omega}$ is plotted as a function of $R_{xx}^5$ at two different $V_g$ in Device 2. The solid lines are linear fittings to the data in the thermal activation regime ($T > 60$ K), which shows that the $R_{NL}^{3\omega}$ scales approximately with the fifth power of the $R_{xx}$.